\documentclass[twocolumn,showpacs,showkeys,preprintnumbers,amsmath,amssymb]{revtex4}

\usepackage{graphicx}
\usepackage{dcolumn}
\usepackage{bm}
\usepackage{longtable}

\begin{document}
\title{Engines at molecular scales}
\author{Raishma Krishnan}
\affiliation{Institute of Physics, Sachivalaya Marg, Bhubaneswar-751005, India}
\author {A. M. Jayannavar}
\email{jayan@iopb.res.in}
\affiliation{Institute of Physics, Sachivalaya Marg, Bhubaneswar-751005, India}
\def\bea{\begin{eqnarray}}
\def\eea{\end{eqnarray}}
\def\be{\begin{equation}}
\parindent=0.7cm
\vspace*{2.5cm}
\begin{abstract}
In recent literature there has been a lot of interest in 
the phenomena of noise induced transport in the absence of 
an average bias occurring in spatially periodic systems 
far from equilibrium. One of the main motivations in this area is to 
understand the mechanism behind the operation of biological 
motors at molecular scale. These molecular motors convert chemical energy 
available during the hydrolysis of ATP into mechanical motion 
to transport cargo and vesicles in living cells with very high 
reliability, adaptability and efficiency in a very noisy environment. 
The basic principle behind such a motion, 
namely the Brownian ratchet principle, has applications 
in nanotechnology as novel nanoparticle separation 
devices. Also, the mechanism of ratchet 
operation finds applications in game theory. Here, we briefly 
focus on the physical concepts underlying the 
constructive role of noise in assisting transport at a molecular 
level. The nature of particle currents, the energetic 
efficiency of these motors, the entropy production in these systems 
and the phenomenon of resonance/coherence are discussed.

\vskip.5cm 
\end{abstract}

\maketitle
 \section{Introduction}
 Noise or fluctuations, that arise either due to the 
 coupling of the system with external incompletely described system or 
 from the bath is traditionally thought of 
 as an unwanted effect. Recently, much work has been done 
 the outcome of which reveal clearly the constructive role 
 of noise in many systems~\cite{julicher,reiman,1amj,astumian,special,
 ptoday,resonance}. Particularly, biological systems provide an important 
 motivation to study the 
 physics of active processes. In the molecular scale, 
 these systems transduce chemical energy obtained from 
 chemical reactions out of equilibrium into mechanical work, 
 generating net motion in a very noisy environment. 
 In terms of magnitude, the particle is acted upon by 
 a noise power of about 
 $8 - 9$ orders of magnitude greater than the chemical power available 
 to drive the motion. Even then the 
 molecular motors, for instance, are able to move and transfer cargo 
 from one point to another and sometimes against the potential gradient.  
 They perform this useful work with high efficiency and reliability 
 even when the environmental conditions are changing all the time. 
 
 Examples of molecular motors include cytoskeletal motor proteins 
 namely kinesin, dynein, etc., which move on the microtubules. 
 Also molecular pumps, for example, sodium or potassium pumps  etc., 
 maintain active transport across membranes against a 
 concentration gradient. What distinguishes these machines 
 from their macroscopic counterparts or heat engines 
 is the fact that they operate 
 in a highly viscous medium which is characterised by low Reynolds 
 number and are subjected to strong thermal fluctuations 
 due to which their motion is stochastic. Hence these 
 motors are termed as Brownian motors 
 or rectifiers. Also, they operate at isothermal conditions. 
 They work by harnessing the force of random motion in the 
 surrounding medium in the absence of a conventional energy 
 source and use it for creating  directed motion. Here we focus 
 mainly on some general physical principles behind such phenomena 
 without going into the details of its biological implications. 

 Any system which is in equilibrium  with a thermal bath at 
 temperature $T$ has the presence of noise in it. Though 
 these thermal noise/fluctuations are ubiquitous, 
 the validity of the second 
 law of thermodynamics forbids the harnessing of noise for 
 useful purposes without spending any energy from the external 
 sources. A Brownian particle executes a random motion 
 in a liquid without any preferential direction. 
 The principle of detailed balance, which essentially means that the rate 
 of forward motion is equal to the rate in 
 the backward direction, forbids current in any preferential direction and 
 hence one cannot extract useful work. In other words, one can extract 
 energy only when the system is driven away from equilibrium. 
 This has been very well demonstrated by 
 Feynman in his `Feynman Lectures on Physics'~\cite{feynmann} by introducing 
 a mechanical ratchet and pawl subjected to thermal fluctuations 
 to demonstrate the impossibility of the  violation of the second law of 
 thermodynamics. Hence building a motor that uses thermal energy 
 from a single heat bath to do mechanical work is not possible.

 \section{Conditions for the effect}
The model to understand such noise 
induced active transport in a fluctuating environment is 
provided by the so called Brownian ratchets.
These are systems with an underlying spatially asymmetric periodic
potential, that exploit the nonequilibrium fluctuations, 
present in the medium, to generate a directed flow. 
The effect of the thermal environment is modeled by considering 
randomly fluctuating force $\xi(t)$ and a concomitant 
viscous (frictional) force with a friction coefficient $\eta$. 
$\eta$ and random noise $\xi(t)$ with $<\xi(t)>=0$ 
are related through the fluctuation-dissipation 
theorem, i.e., $<\xi(t)\xi(s)>=2\eta k_BT \delta(t-s)$. 
Physical models like flashing ratchets, 
rocking ratchets, time asymmetric ratchets, inhomogeneous (frictional) 
ratchets etc., have been proposed to achieve essential 
nonequilibrium conditions for net motion in a periodic 
system. As long as the system is left alone and 
remains in thermodynamic equilibrium
particles in a ratchet cannot diffuse in any 
preferential direction, in spite of the spatial 
asymmetry in the potential. But in the presence of 
additive or multiplicative nonequilibrium fluctuations 
the particles in general start to move in one direction 
as the principle of detailed balance does not hold in this case. 
Thus both nonequilibrium fluctuations 
and spatial or temporal asymmetry in potential 
conspire to generate a unidirectional flow in the absence of bias. 
In the following we briefly discuss the essential ideas 
behind some of the ratchet models.

\section{Different types of ratchets}
\subsection{Flashing ratchets}\label{fl}
This is a simple model that closely resembles the mode of 
operation of protein motors. In this model the periodic potential 
is allowed to fluctuate 
with finite time correlation between two states 
characterised by different barrier heights. For example, 
the overdamped Brownian particles are subjected to two 
potential states periodically, i.e., $V_{on}$ which 
corresponds to an asymmetric saw tooth like 
potential for a time $\tau_{on}$
and  $V_{off}$ which 
corresponds to zero potential (flat) state for a time interval $\tau_{off}$ 
as is shown in Fig.~\ref{flashing}. 
\begin{figure}[htp!]
 \begin{center}
\input{epsf}
\hskip15cm \epsfxsize=3in \epsfbox{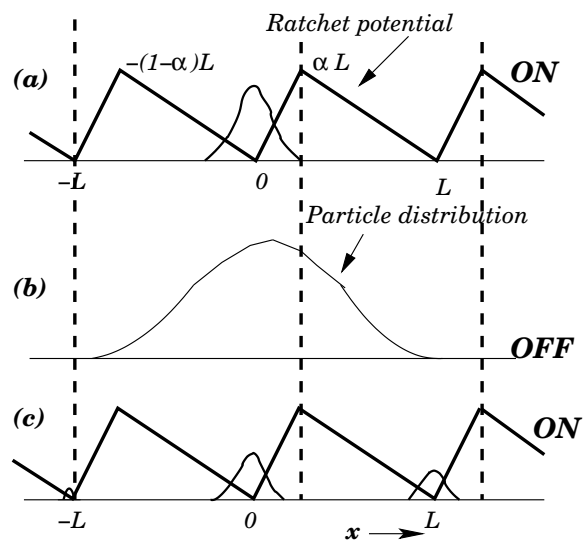}
 \caption{Flashing ratchet model. Here $L$ is 
 the period of the potential and $\alpha$ is the 
 asymmetry parameter.} \label{flashing}
 \end{center}
 \end{figure}
During the period when the potential is {\it{on}} the particles 
will slide down the potential slope to the bottom of 
the potential minima due to which there is a peak in 
the probability density of particles at these minima. 
Switching the potential {\it{off}} allows the particles to 
diffuse freely and the density of particles spread into a Gaussian 
curve centered around the minima as shown in Fig.~\ref{flashing}b. 
At the end of $\tau_{off}$ the potential is again put back to the 
{\it{on}} state for an interval $\tau_{on}$ and the particles will again 
slide down along the direction of local force to the nearest minima, 
Fig.~\ref{flashing}c. This process is continued indefinitely 
resulting in a net current in one direction because of 
spatial asymmetry in the potential within a period.

The main point to be taken care of is the time 
interval between switching {\it{on}} and {\it{off}} periods of the potential. 
If $\tau_{off}$ is adjusted such that by the end of $\tau_{off}$ 
the diffusive motion takes the particle out of the 
earlier existing potential minima in the steeper slope 
(smaller distance) direction of the saw tooth 
potential but fails to do so in as much proportion in the gentler slope 
(larger distance) direction, then in the next {\it{on}} 
interval of the potential 
the particles will slide along the gentler slope 
to the adjacent minimum of the saw tooth potential. 
This process of sequential flipping of the potential 
between {\it{on}} and {\it{off}} states  is continued and in the long time 
limit one gets a net flow of current to the  
right side. The system is supplied with the required 
energy to flip the potential states externally thereby 
rendering the system nonequilibrium. It is to be emphasized that 
thermal fluctuations are necessary for the working of flashing ratchets.
Moreover, no macroscopic bias is applied to the system.

\subsubsection{Equivalence to Carnot engine}
The flashing ratchet model where the potential is flipped 
between {\it{on}} and {\it{off}} states is analogous to the case where 
the particle is coupled to two temperature baths.
From statistical mechanics it is well known that the probability 
to cross a barrier is governed by the factor $\exp-(V/k_BT)$ where $V$ is the 
potential barrier height. When the temperature $T$ is 
large the average kinetic energy of the 
particle is large and so is the violent thermal fluctuations. 
Due to this the particle hardly feels the presence of potential in 
comparison with the thermal noise and hence the probability to cross 
the barriers is large. This is equivalent to the 
particle being in the {\it{off}} state of the flashing 
ratchet discussed above. In the opposite case when 
the temperature of the medium is small, 
the average kinetic energy and the thermal fluctuations are 
small and the particles will feel the presence of the asymmetric 
potential. This case is akin to the {\it{on}} state in the flashing 
ratchet. Thus the systematic coupling of the particle randomly 
to two temperature baths is equivalent to flashing ratchet model 
where the potential is switched {\it{on}} and {\it{off}} and in the 
long time limit one gets unidirectional current. In this spirit the ratchet 
system has direct equivalence to a Carnot engine which 
extracts work by making use of two thermal baths held at 
different temperatures. 

The relevant system variables for the ratchets are 
temperature $T(t)$ and the position $x(t)$ while that for a 
Carnot engine are pressure and volume. However, there 
exists qualitative differences between the two. 
For the case of ratchets, after one 
periodic variation in time of $T(t)$ or one temperature cycle the 
particle may or maynot come to the same position or in 
otherwords there is no synchronization between the relevant 
system variables. But for a Carnot engine there is a complete 
synchronization between the relevant  system variables along the 
cycle. The ideal Carnot engine moreover will not be 
simultaneously in contact with two temperature baths 
assuring the reversible mode of operation. In contrast,
ratchets or molecular motors  work in an intrinsically irreversible 
mode of operation with a very low efficiency. It has to be emphasised that 
Carnot engine gives high efficiency only in the quasi-static 
mode of operation  and 
though there is net work done by the engine the net power delivered 
in the cycle is zero. For the case of molecular motors we may 
get a higher efficiency in the nonadiabatic regime 
(i.e. by increasing the frequency of oscillation or cycling) as compared to 
the values obtained in the adiabatic 
or quasi-static regime of operation. This behaviour is quite 
contrary to the case of reversible macroscopic heat engines. The 
distinguishing factor of a Brownian motor is that noise plays 
a dominant role and that noise may facilitate energy 
transduction leading to high efficiency of these molecular 
engines which is counterintuitive~\cite{danenergetics-adia}. 

In these molecular engines noise and associated dispersion 
of particles are no longer thought of as a hindrance, 
but are instead incorporated as a part of the design. Moreover, an 
efficient microscopic engine is not necessarily the microscopic 
equivalent of an efficient macroscopic engine.  

\subsubsection{Current reversals and mass separation devices}
With a  judicious choice of asymmetric potential the current reverses 
its sign as a function of suitable system parameter. 
This phenomenon is called current reversal. Thus, Brownian particles 
with different friction coefficients, masses or charges move in 
opposite directions and hence they can be readily separated. 
This is a new modern method of separating particles at nanometer scale. 
\begin{figure}[htp!]
 \begin{center}
\input{epsf}
\hskip15cm \epsfxsize=2in \epsfbox{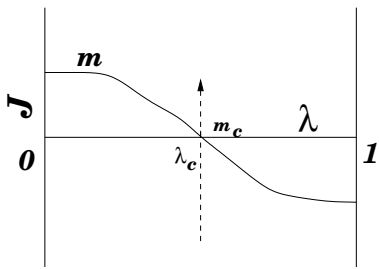}
 \caption{Figure to illustrate current reversal.}\label{mass}
 \end{center}
 \end{figure}
To illustrate the phenomenon of current reversal consider a 
potential of the form $V_\lambda(x,t)=\lambda V_2(x,t) + (1-\lambda) V_1(x,t)$ 
where $V_1(x,t)$ and $V_2(x,t)$ are two fluctuating ratchet 
potentials such that the unidirectional 
flow of currents in $V_1(x,t)$ and $V_2(x,t)$  are in opposite 
directions in the presence of nonequilibrium 
fluctuations and $\lambda$ is a parameter between $0$ and $1$. 
If $V_2(x,t)$ is a mirror reflection of $V_1(x,t)$, the fluctuating 
ratchet potential shown in Fig.~\ref{flashing}, then under the influence 
of $V_2(x,t)$ alone current flows in the negative direction. 
The current has a smooth dependence on mass and the higher the mass 
the lower is the current. A particle of mass $m$ will 
move in respective directions depending on the potential 
to which it is subjected to. Suppose that the particle is 
subjected to the potential $V_\lambda(x)$ which is a combination 
of $V_1(x,t)$ and $V_2(x,t)$. The plausible curve for unidirectional 
current as a function of $\lambda$ in shown Fig.~\ref{mass}. When $\lambda$ 
is zero there will be contribution only from $V_1(x,t)$ 
and one gets a current which is positive. 
When $\lambda=1$ the contribution to the current will be 
from $V_2(x,t)$ and hence the flow will be in the 
opposite direction. Thus by continuously deforming one 
potential into another, i.e., $V_1(x,t)$ into $V_2(x,t)$ 
there must exist an intermediate potential with the property 
that the particle current is zero at some finite value of the 
parameter $\lambda_c$. Hence there is a critical 
$\lambda_c$ at which the current curve 
passes with a finite slope through this zero point 
thereby implying the existence of current inversion 
as a function of $\lambda$. 

Once a current inversion upon variation of 
one parameter of the model is established an inversion 
upon variation of any other parameter can be inferred along 
the same line of reasoning. Suppose we fix a point, 
say, the value of $\lambda= \lambda_c$ at which 
the current for a particular mass, $m_c$, is zero. Now if we 
vary the mass around the value $m_c$ one would see that the current as  
a function of mass will smoothly go through this 
zero point at $m=m_c$ with a finite slope. This 
means that a current reversal is obtained as a function of mass. 
That is, particles of mass greater than $m_c$ and that with mass 
less than $m_c$ get separated in opposite directions.

The phenomenon of current reversal can play a major role 
in separation devices. This method of particle separation 
has many features far superior 
than the existing methods like electrophoresis, centrifugation, 
chromatography etc., which rely on the motion caused by 
long range gradients. In these methods the thermal 
noise inturn degrades 
the quality of separation due to the diffusive broadening of the bands.   
It is also possible to get multiple
current reversals by properly choosing the potential 
as a function of system parameters. By multiple 
reversals one can separate blocks of particles of different
masses with parameters within a characteristic window. 
It may be noted that for two dimensional ratchets, particles 
of different masses get separated in different directions.     

\subsubsection{Parrondo's paradox}
The concept of Brownian ratchets, where there 
is rectification of fluctuations to give unidirectional 
current also has its extensions to game 
theory opening up a new area of paradoxical gambling games
under the subject of Parrondo's paradoxes~\cite{parrondoparadox-review}. 
These games can be thought of as a discrete time 
version of flashing Brownian ratchets with the interesting 
consequence that by randomly switching between two losing games one 
tends to win.  
\begin{figure}[htp!]
 \begin{center}
\input{epsf}
\hskip15cm \epsfxsize=3.4in \epsfbox{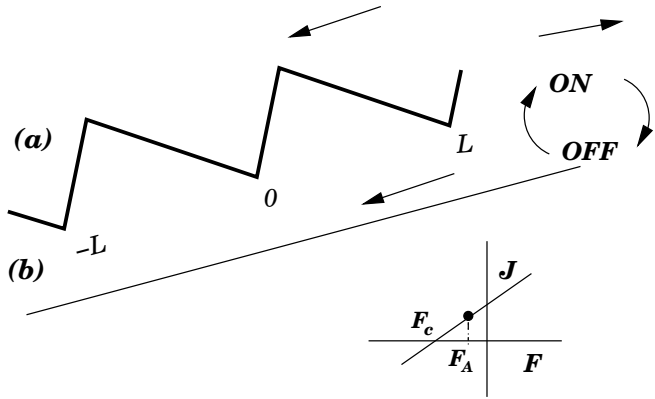}
 \caption{Illustration of Parrando's paradox. (a) and (b) corresponds 
 to the {\it{on}} and {\it{off}} potential states in the 
 presence of bias field.} \label{tilt}
 \end{center}
 \end{figure}
To begin with, consider a flashing ratchet 
model in the absence of any external force 
as discussed in Sec.~\ref{fl} with a current flow 
towards the right, Fig.~\ref{flashing}, or in the positive direction 
when the potential is fluctuating 
between the {\it{on}} and {\it{off}} states. To this a constant tilting  
force $F$, as in  Fig.~\ref{tilt}, is applied opposite 
to the direction of current in the ratchet. 
Then, as expected, the current ($J$) in the positive direction decreases 
and beyond a particular value called the stopping force $F_c$ 
the current crosses over to the negative direction. This is 
illustrated in the $J$ vs $F$ plot in Fig.~\ref{tilt}. 
Let us consider a particular value of force say 
$F_A$ in the $J$ vs $F$ plot. 
The corresponding flashing ratchet profile is as in the figure. 
In the presence of a bias force $F_A$, the current flows 
in the negative direction as expected in both the {\it{on}} and 
{\it{off}} states of the potential in the absence of flipping. 
But flipping between the {\it{on}} and {\it{off}} states 
randomly for a long time will result in 
a current in the positive direction. Thus at the point $F_A$ one gets 
current in the positive direction though when considered 
separetely ( i.e., when the potential is as shown either 
in Fig.~\ref{tilt}a or in Fig.~\ref{tilt}b) the flow of current is in 
the negative direction.

The current in the negative direction for the two different potentials is 
analogous to a losing game and thus alternating randomly 
between these two losing games one has a finite probability to win 
(current in the positive direction). 
The game reveals the fact that the outcome of the alternation 
of two stochastic dynamics can significantly be different 
from each separate one. Parrando's paradox, where 
one basically converts two losing strategies into a winning one, 
has numerous applications in the field of economics, sociology and 
many other interdisciplinary areas. An example is the formation of spatial 
patterns in spatially extended systems by alternation 
between two dynamics which in turn is absent in the presence of 
only a single dynamics. 

\subsection{Rocking ratchets}\label{rocking}
In the ratchet model we had considered before namely, flashing 
ratchet, the potential fluctuates between {\it{on}} and {\it{off}} states. 
Another type of ratchet corresponds to rocking ratchets 
where given an asymmetric potential one applies a random 
time varying force with mean zero. Due to the anisotropy 
of the potential (a special case is shown in Fig.~\ref{rock}a), 
when a force having same magnitude but different signs $+F$ and $-F$ 
are applied, the motion of the particle on the average 
will be along positive and negative direction respectively. However, 
particles will have to overcome only the smaller barriers 
along the direction of their average motion in the presence 
of a positive force as opposed to the case when the force 
is negative with the same magnitude. 
\begin{figure}[htp!]
 \begin{center}
\input{epsf}
\hskip15cm \epsfxsize=2.5in \epsfbox{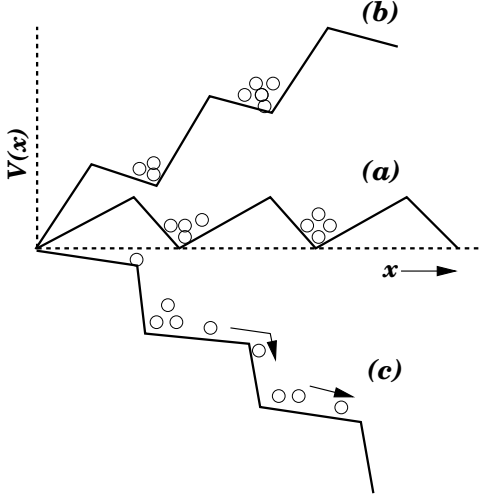}
 \caption{Rocking ratchet model.} \label{rock}
 \end{center}
 \end{figure}
We consider a case where the average slope of 
the saw tooth potential, Fig.~\ref{rock}a, is changed 
in time either slowly or abruptly with a finite maximum 
value on either side of the zero slope line ensuring that the 
time average of the force acting on the particle due 
to rocking is zero (Fig.~\ref{rock}b and c). 
The rocking or changing of slopes can be done either periodically or 
randomly in time. 

At very low temperature when the particles do not have enough energy to 
overcome the barrier the particles get trapped in the 
minima of the potential. A special case of rocking force, $+F$ and $-F$, 
imposed on the ratchet potential is shown in Fig.~\ref{rock}b 
and ~\ref{rock}c respectively.  For the case when force is negative, 
Fig.~\ref{rock}b, 
the particle remains trapped in one of the trenches 
whereas when the force is positive, Fig.~\ref{rock}c,
the particle is capable of running down the potential hill. 
With such a geometrical construction one can notice that the 
current when the force 
is $+|F|$ is not equal and opposite to the case when force
is  $-|F|$. In otherwords, $J(|F|) \neq - J(-|F|)$ remains 
valid even for finite temperatures. Thus system acts as a nonlinear 
rectifier in the presence of zero average periodic or random force.

Unlike in the case of flashing ratchets, the direction of 
current for the rocking ratchet is in the direction of the steeper slope 
and this mechanism of rocking is equivalent to generating 
dc current in semiconductor pn junctions under an applied ac bias.  

\subsection{Inhomogeneous ratchets}
There is yet another type of ratchet, namely, 
frictional ratchets~\cite{1amj,enslaving} 
which unlike the ones discussed above gives unidirectional 
current even in the presence of spatially periodic symmetric 
potential, $V(x)$ but in the presence of space dependent diffusion coefficient 
$D(x)=k_BT(x)/\eta(x)$.  
The space dependence of diffusion coefficient could arise either 
due to space dependent temperature $T(x)$ or space dependent 
friction coefficient $\eta(x)$. Such inhomogeneous systems 
are common in nature. For example, particles diffusing close 
to surface have space dependent friction coefficient. 
The molecular motor proteins are believed to 
be moving close along the microtubules and therefore experience 
space dependent mobility. Semiconductor systems and superlattice 
structures also have space dependent friction coefficient.  
\begin{figure}[htp!]
 \begin{center}
\input{epsf}
\hskip15cm \epsfxsize=3in \epsfbox{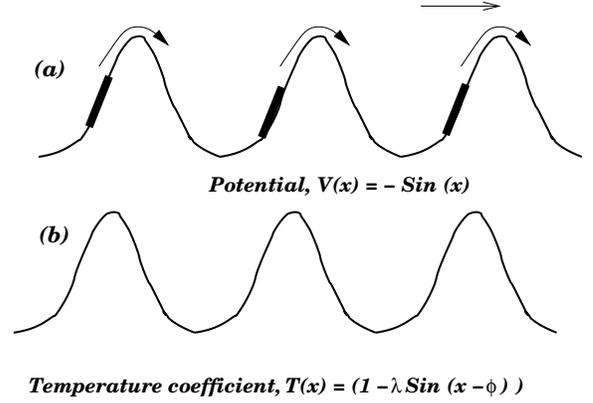}
 \caption{Illustration of an inhomogeneous ratchet. The temperature 
 and potential profiles are depicted in the figure.} \label{frictional}
 \end{center}
 \end{figure}
The peculiarity of this type of ratchets is that the 
system dissipates energy during its time evolution 
differently at different places due to the space dependent 
diffusion coefficient arising from variation in temperature $T(x)$ 
which in turn implies that the system is out of equilibrium.  
The only criterion that has to be satisfied here is 
that both the potential $V(x)$ as well as temperture 
$T(x)$ has to be periodic and 
should be separated by a phase difference other than $0$ 
and $\pi$ with respect to the potential as shown in Fig.~\ref{frictional}. 
A similar effect of 
variation in temperature can also be obtained if we have a medium with 
space dependent friction coefficient $\eta(x)$ in the presence of 
external noise. 
 The noise being externally imposed the system 
always absorbs energy due to the absence of a concomitant loss factor. 
At the regions where the friction 
coefficient is high the overdamped particle stays for a longer time 
due to which the possibility of absorption of 
energy from the external noise in that region is correspondingly high. 
This leads to an increase in the local temperature in these regions.
Thus system with  space dependent friction at 
constant temperature in the presence of external parametric 
noise  is equivalent to a system with a space  dependent 
temperature field.  

To illustrate the net unidirectional transport in 
these systems consider a periodic potential $V(x)$ as shown  
in Fig.~\ref{frictional}a. Also consider a space dependent 
temperature profile with same periodicity as that of 
potential but shifted by a phase difference $\phi$ as 
shown in Fig.~\ref{frictional}b. In Fig.~\ref{frictional}a 
the darkened regions specify regions of higher temperature 
corresponding to the regions where the temperature 
profile has peaks Fig.~\ref{frictional}b. The particle in 
the darkened regions (high temperature regions) on 
the average gains more energy as compared to other regions. As a 
consequence the particle in any potential minima will  find it 
easier to cross the peak of the potential and go over to the 
right side than to the left side. Hence current 
in the right side is assured. The magnitude as well as 
the direction of current depends on the phase difference $\phi$.

It may be noted that unidirectional motion in inhomogeneous systems 
arise as a corrollary to the well known Landauer's blow 
torch principle~\cite{landauer}. This principle states that the behaviour 
of nonequilibrium systems will depend sensitively on the specific details 
of its kinetics, even on pathways that traverse infrequently 
occupied kinetic states far from the stable state. In otherwords, 
the stability criteria which examine only the immediate vicinity 
of a locally stable state are inadequate to assess the relative 
stability of states in nonequilibrium systems. In contrast, 
microscopically reversible systems can well be characterised by criteria 
that depend only on the local neighbourhood of the equilibrium state.
\section{Energetics of Brownian motors}
As mentioned above, ratchets extract energy from random 
fluctuations and generate currents or ordered motion. 
In this sense they can be considered as information 
engines analogous to Maxwell's demon which extract 
work out of bath at the expense of an overall 
increase in entropy. The usefulness of any engine lie in the 
extent of work that can be efficiently extracted out of it. 
The molecular motors in living cells are found to be very 
efficient in their noisy environment. 
In all the ratchet models that we have 
discussed above no useful work has been performed. This 
is because particles moving in a periodic potential system 
ends up with the same potential energy 
even after crossing over the adjacent potential minimum. 
There is no extra energy stored 
in the particle which can be usefully expended when needed. 
To have an engine out of a ratchet it is necessary to use 
its systematic motion to store potential energy 
which inturn is achieved if a ratchet lifts a load.  
\begin{figure}[htp!]
 \begin{center}
\input{epsf}
\hskip15cm \epsfxsize=3in \epsfbox{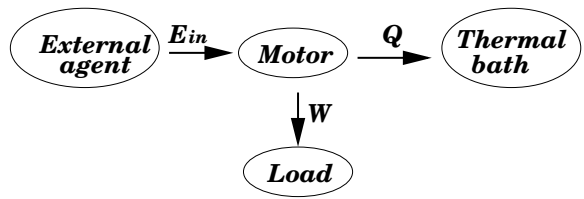}
 \caption{Schematic figure of the energy flow in a Brownian motor.}
 \label{energetics}
 \end{center}
 \end{figure}
Thus for the ratchet to perform work a small force 
called load (Fig.~\ref{energetics}) has to be applied opposite to the 
direction of current in the ratchet. Then the particles 
will keep on moving, on the average, against the force or 
load performing work. Part of the input 
energy, $E_{in}$, coming from the source of nonequilibrium 
is converted into mechanical energy related to the load. 

A general framework has been developed wherein 
the compatibility between the Langevin or Fokker-Planck 
formalisms, used to discuss stochastic processes, 
and the laws of thermodynamics, which characterize the 
thermal and mechanical behaviour of macroscopic systems, 
have been established~\cite{sekimotoparrondo}. The concept of heat 
on mesoscopic scales has been defined in terms of Langevin 
dynamics and the essential point 
behind this formalism is that the heat transferred to the system 
is nothing but the microscopic work done by both the 
frictional and random force in the Langevin equation 
(i.e., work done by the bath on the system). This is also consistent 
with the fact that we cannot control all the details of energy transfer 
which inturn leads to the concept of heat (via stochastic dynamics) 
as a form of energy flow. The subject of the energetics of Brownian 
motors has developed into an entire subfield on its own right. 
Fig.~\ref{energetics} represents the energy flow between the isothermal 
Brownian motor and its surroundings. The first and the 
second law of thermodynamics is now given as
\begin{eqnarray}
E_{in} &=& Q+ W  \label{one}\\
 \Delta S_{agent} + \frac{Q}{T} &=& S_{prod} \geq 0 \label{two}
\end{eqnarray}
In the above equation $E_{in}$ is the input energy into the system from 
the external agent, $Q$, the heat dissipated to the bath, 
and $W$ the work done. All these quantities can be defined 
for each microscopic realization of the motion of Brownian 
particle or motor. Eqns.~\ref{one}  and ~\ref{two} correspond to the 
first and second law of thermodynamics. 
Again, $\Delta S_{agent}$ is the change in entropy of the external 
agent, $Q/T$ is the entropy given to the bath and $S_{prod}$ is the 
total entropy production of the universe. $T$ corresponds to the absolute 
temperature. Magnitudes of all the physical quantities are 
taken over a cycle or per unit time in the stationary regime. 
In this regime the entropy and the internal energy 
of the motor (system) which are the state variables does not 
change. The formal expressions for all the above 
mentioned physical quantities are known in terms of 
the probability distribution of the particles.  

Using this framework of stochastic energetics 
one can readily calculate various physical 
quantities like efficiency of energy conversion ($\eta = W/E_{in}$),
energy dissipation (hysteresis loss), entropy production
~\cite{rk-dan-amj-physica}, 
input energy, work etc. The important point to be noted here is that an 
analysis of fluctuations, which is completely ignored 
in the working of heat engines at larger scale, is essential 
for the calculation of efficiency of ratchet systems at 
the molecular scale. The efficiency of the Brownian 
motors is sensitively dependent on system parameters. 

The study of the  efficiency of energy transduction 
by different types of ratchet models show the ratchets to 
have very low efficiency. The observed efficiency values of the 
several ratchet models like flashing ratchets, rocking ratchets 
etc., are found to fall in the subpercentage regime $(< .01)$.
This is due to the fact that every time the potential changes the 
particle distribution also changes and tries to 
adapt to the changing environmental conditions. This leads to 
an inevitable loss in the medium or in other words the mode of 
operation of the ratchets is intrinsically irreversible. As a 
consequence the unattainability of Carnot efficieny 
in Brownian heat engines has been emphasized in literature. 
Currently, the notion of reversible ratchets where the energy 
dissipation or entropy production is almost zero are being pursued. 
These reversible ratchets are sometimes termed as adiabatic pumps wherein 
transport of particles with zero entropy production is generated by 
cyclic adiabatic variations of atleast two parameters 
of the periodic potential (which are out of phase in time) 
in the absence of bias~\cite{sekimotoparrondo}. 

The energetic efficiency of a ratchet is not 
an intrinsic property of the device and it depends on 
the characteristics of the imposed external load. 
By a judicial choice of the external load one 
can improve the efficiency considerably. Recently~\cite{tsong} a flashing 
rachet model has been developed, wherein with two asymmetric double-well 
periodic-potential- states displaced by half a period a high efficiency 
has been achieved due to the blocking of particle 
motion in the opposite direction to that of the net average current.  
Such flashing ratchet models were found to be highly 
efficient with efficiency
an order of magnitude higher than in the earlier models.
The basic idea behind this enhanced efficiency is that even for 
diffusive Brownian motion the choice of appropriate 
potential profile ensures suppression of backward motion 
and hence a reduction in the accompanying dissipation.

We have studied the motion of a particle in a 
new class of rocking ratchets rocked purposefully as 
to favour current in one direction but to suppress 
motion in the opposite direction.  
This is accomplished by applying a temporally 
asymmetric but unbiased periodic forcings.
It may be noted that in this type of ratchets a larger 
force field is applied for a short time interval 
of the period in the forward direction as compared to 
a smaller force for a longer time interval in the other 
direction. The intervals are so chosen that the net 
external force or bias acting on the particle over 
a period is zero. In these new class of temporally 
asymmetric driven ratchets one gets unidirectional 
current even in the presence of spatially symmetric 
potential~\cite{mangalmillonas}.  

At low temperatures when $k_BT$ is much 
less than $V_0$, the modulation amplitude of the 
periodic symmetric potential, significant current arises only 
when the bias field is greater than a critical field 
$F_c$, the value of which should be greater 
than $V_0$~\cite{riskin}. If the bias field is 
less than $V_0$, the particle will feel the barriers and hence 
current flux in the negative direction is very 
small or there is blocking of current. A significant current 
flux  in the positive direction arises only when 
the temporally asymmetric bias force field in that direction is greater 
than $V_0$. When this condition, of bias field being less than 
$V_0$ in the backward direction and being greater than 
$V_0$ in the forward direction, is satisfied the barriers for 
motion in the forward direction disappears and one 
gets unidirectional current. Interestingly, such choice 
of forcings help in obtaining rectified currents with 
high efficiency of the order of $50\%$  without fine 
tuning of the physical parameters. This efficiency is several 
orders of magnitude larger than the obtained efficiency in several 
other ratchet models. Moreover, the range of parameters of operation 
of such ratchets is quite wide sustaining large loads. 

\subsection{Currents, Stochastic resonance and Coherence}
The noise induced currents in most of the ratchet systems 
exhibits a peak with respect to noise strength 
and other physical parameters. Such peaking behaviour 
is expected when system exhibits a resonance phenomena. 
Infact, some recent studies have tried to reveal the 
relation between two unrelated phenomena, namely stochastic resonance 
and Brownian ratchets in a formal way through the consideration of 
Fokker-Planck equations. We have analysed this issue 
by using the method of stochastic energetics in several 
classes of adiabatically rocked ratchets. 

The resonance behaviour can be well characterised by 
the behaviour of input energy. It is expected that at resonance 
the system extracts maximum energy from the external source and 
hence in the time periodic stationary state this energy is 
dissipated to the environment (hysteresis loss). Our studies show 
the input energy to have a monotonous behaviour as a function of noise 
strength. Thus the resonance like feature observed in the nature 
of currents as a function of noise strength is not related to 
the intrinsic resonance in the dynamics of the particle 
with the external ac drive. The above observations are valid 
only for a class of adiabatically rocked ratchets.  

The presence of net currents (ordered motion) in the ratchets increases the 
amount of known information about the system than otherwise. This 
extra bit of information comes from the negentropy or the physical 
information supplied by the external nonequilibrium bath. The amount 
of information that is  transferred by the nonequilibrium bath is quantified 
in terms of algorithmic complexity of the position of Brownian particle. 
It has been argued that the algorithmic complexity or Kolmogorov 
information entropy is maximum when the current is maximum~\cite{san}. 
Since the currents are generated at the expense of entropy we 
naively expect the maxima in current to be related to the maxima 
in the overall entropy production as a function of noise strength. 
However, we have shown that the total entropy production 
does not extremize at the same parameter 
value at which the current exhibits a maximum~\cite{rk-dan-amj-physica}.
\begin{figure}[htp!]
\input{epsf}
\hskip15cm \epsfxsize=3.7in \epsfbox{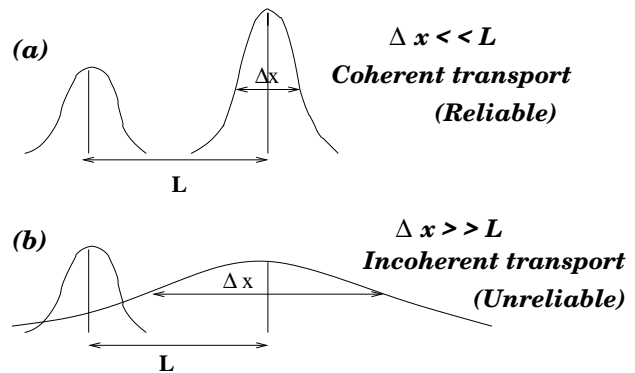}
 \caption{Evolution of particle distribution for a given time interval 
 is depicted for two separate cases of particle transport.} \label{peclet}
 \end{figure}
The fact that noise strength at which the maxima seen in both 
current and entropy production do not coincide may be 
related to the quality of the current or the coherence 
in transport. Noise induced currents are always 
accompanied by a dispersion or diffusion. 
When the diffusion is large, $\Delta x >>L$ with 
$\Delta x$ being the diffusive spread when the mean 
position of the particle is shifted by a distance $L$ which is 
the period of the potential, then the quality of transport degrades 
and the coherence in the unidirectional motion is lost. The 
coherent transport (optimal transport) refers to the case of 
large mean velocity at fairly small diffusion (see Fig.~\ref{peclet}). 
This is in turn quantified by a dimensionless 
Pe\'{c}let number which is the ratio of current to the 
diffusion constant~\cite{Pec}. From  
Fig.~\ref{peclet} we see that in both the cases 
there is a noise induced current of the same magnitude, but transport in 
Fig.~\ref{peclet}a is more coherent due to the fact 
that particles are reliably transferred from one 
point to the other due to the less diffusive spread. 
In Fig.~\ref{peclet}b though there is a shift in 
the peak (or there is current of same magnitude) due to the large diffusive 
spread the probability that the particle is delivered 
to the desired region is less. There is a finite probability 
that the particle may still be around the region 
where it has started.    

For a given magnitude of current the transport may be 
coherent or incoherent. Thus analysis 
of the relation between current and the entropy production requires 
not only the magnitude of current but also the quality of 
transport. These studies are expected to reveal a deep connection between 
efficiency, quality of transport, entropy and information. 

\section{Conclusions}
To summarize, we have given a qualitative picture of the constructive role of 
noise in nonequilibrium systems. This area has attracted great 
interest from diverse areas of science and technology. We have presented a  
brief description of the different ratchet models or Brownian motors 
and also the method of stochastic energetics that was developed in 
order to understand the energetics in such systems with special 
emphasis on our work and results. 

In our discussion so far we had restricted only to the case of 
isolated Brownian motors. Infact, the cooperativity among Brownian motors 
has far reaching consequences~\cite{reiman}. The coupling 
among these motors can 
lead to a marked increase in  the efficiency of energy transduction 
as well as the magnitude of macroscopic current. Cooperative motors 
also exhibit other fascinating phenomena such as phase transitions, 
normal to anomalous hysteretic behaviour, absolute negative mobility etc. 
The parallel development in mesoscopic systems has led to the 
discovery of quantum ratchets. These quantum ratchets make use of 
quantum effects such as tunneling and wave interference effects. 
Such electron ratchets can not only be used to generate particle 
current but also to pump heat in the reversible mode of operation. 

To conclude, we have shown that molecular motors work as engines at 
molecular scale. These microscopic engines are not the microscopic equivalent 
of the efficient macroscopic engines that we come across in 
our daily life. Noise play an inherent constructive role in the 
mode of operation of these engines. Thus noise is not a nuisance 
but rather an inseparable part of the design of the engine operation. 
In some cases increase in noise strength is found to even enhance 
the efficiency of these engines.
The operation of these engines are done by the engines themselves and 
they are out of equilibrium. As such there are no general principles 
that determines the mode of optimal efficiency of these engines. Further 
studies in this interdisciplinary area could lead to a better 
knowledge of the functioning of these biological motors in 
living cells and also in the creation of efficient 
man made nanoscale machines. Such studies could also bear 
important consequences in understanding the fundamental issues 
in nonequilibrium statistical mechanics.


 \end{document}